# The Lowell Database Research Self Assessment

By Authors [*]

June 2003

## 1 Summary

A group of senior database researchers gathers every few years to assess the state of database research and to point out problem areas that deserve additional focus. This report summarizes the discussion and conclusions of the sixth ad-hoc meeting held May 4-6, 2003 in Lowell, Mass. It observes that information management continues to be a critical component of most complex software systems. It recommends that database researchers increase focus on: integration of text, data, code, and streams; fusion of information from heterogeneous data sources; reasoning about uncertain data; unsupervised data mining for interesting correlations; information privacy; and self-adaptation and repair.

## 2 Introduction

Some database researchers have gathered every few years to assess the state of database research and to recommend problem areas that deserve additional focus. This report follows a number of earlier reports with similar goals, including: Laguna Beach, California in 1989 [1], Palo Alto, California ("Lagunita") in 1990 [2] and 1995 [3], Cambridge, Massachusetts in 1996 [4], and Asilomar, California in 1998 [5]. Continuing this tradition, 25 senior database researchers representing a broad cross section of the field in terms of research interests, affiliations, and geography, gathered in Lowell, Mass. in early May, 2003 for two days of intensive discussion on where the database field is and where it should be going. Several important observations came out of this meeting.

Our community focuses on information storage, organization, management, and access and it is driven by new applications, technology trends, new synergies with related fields, and innovation within the field itself. The nature and sources of information are changing. Everyone is aware that the Internet, the Web, science, and eCommerce are enormous sources of information and information-processing demands. Another big source is coming soon: cheap microsensor technology that will enable most things to report their state in real time. This information will support applications whose main purpose is to monitor the object's state or location. The world of sensor-information processing will raise many of the most interesting database issues in a new environment, with a new set of constraints and opportunities.

In the area of applications, the Internet is currently the main driving force, particularly by enabling "cross enterprise" applications. Historically, applications were intra-enterprise and could be specified and optimized entirely within one administrative domain. However, most enterprises are interested in interacting with their suppliers and customers to share information and provide better customer support. Such applications are fundamentally cross-enterprise and require stronger facilities for security and information integration. They generate new issues for the Database Management System (DBMS) community to deal with.

A second application area of growing importance is the sciences — notably the physical sciences, biological sciences, health sciences, and engineering — which are generating large and complex data sets that need more advanced database support than current products provide. They also need information integration mechanisms. In addition, they need help with managing the pipeline of data products produced by data analysis, storing and querying "ordered" data (e.g., time series, image analysis, computational meshes, and geographic information), and integrating with the world-wide data grid.

In addition to these new information-management challenges, we face major changes in the traditional DBMS topics such as data models, access methods, query processing algorithms, concurrency control, recovery, query languages, and user interfaces to DBMSs. These topics have been well studied in the past. However, technology keeps changing the rules. For example, disks and RAM are getting much larger and much cheaper per bit of storage. Access times and bandwidths are improving too, but they are not improving as rapidly as capacity and cost. These changing ratios require us to reassess storage management and query-processing algorithms. In addition, processor caches have exploded in size and levels, requiring DBMS algorithms to be cache-aware. These are but two examples of technological change inducing a reassessment of previous algorithms in light of the new state of affairs.

---

[*] Attendees at the Lowell Workshop were: Serge Abiteboul, Rakesh Agrawal, Phil Bernstein, Mike Carey, Stefano Ceri, Bruce Croft, David DeWitt, Mike Franklin, Hector Garcia Molina, Dieter Gawlick, Jim Gray, Laura Haas, Alon Halevy, Joe Hellerstein, Yannis Ioannidis, Martin Kersten, Michael Pazzani, Mike Lesk, David Maier, Jeff Naughton, Hans Schek, Timos Sellis, Avi Silberschatz, Mike Stonebraker, Rick Snodgrass, Jeff Ullman, Gerhard Weikum, Jennifer Widom, and Stan Zdonik. Slides and some detailed notes from the event are at http://research.microsoft.com/~gray/lowell/.

Another driver of database research is the maturation of related technologies. For example, over the past decade, data-mining technology has become an important component of database systems. Web search engines have made information retrieval a commodity that needs to be integrated with classical database search techniques. Many areas of artificial intelligence are producing components that could be combined with database techniques; for example these components allow us to handle speech, natural language, reasoning with uncertainty, and machine learning.

Participants noted that it is a popular undertaking these days to propose "grand challenges" for various fields of computer science. Each grand challenge is a problem that cannot be solved easily, and is intended as a "call to action" for a given field, such as The Information Utility [5] and Building Systems With Billions of Parts [6]. We all agreed that we could define more grand challenges. In fact, we discussed a few, notably the personal information manager — a database that could store, organize and provide access to all of a person's digitally-encoded information for a lifetime. But in the end, we decided that focusing on a single grand challenge was inappropriate, since information management technology is a critical component in most, if not all, of the proposed computer-science grand challenges. Moreover, many of those information-management challenges are well beyond the state of the art. The existing grand challenges are a full-employment act for the database community — we decided not to add any more.

During the two days, we noted many new applications, technology trends, and synergies with related fields that affect information management. In aggregate, these issues require a new information-management infrastructure that is different from the one used today. Hence, Section 2 surveys the components of this infrastructure. Section 3 presents a short discussion of the topics that generated controversy during the meeting, and a statement of next steps that can be taken to move the new information management infrastructure closer to reality.

## 3    Next Generation Infrastructure

This section discusses the various infrastructure components that require new solutions or are novel in some other way.

### 3.1    Integration of Text, Data, Code, and Streams

The DBMS field has always focused on capturing, organizing, storing, analyzing, and retrieving structured data. Until recently, there was limited interest in extending a DBMS to also manage text, temporal, spatial, sound, image, or video data. However, the Web has clearly demonstrated the importance of these more sophisticated data types. The general problem is that as systems add capabilities, it is hard to make these additions "cleanly." Rather, there is a tendency to do the minimum necessary to have the most important of the desired new features. As a result, these extensions tend create "second-class citizens" — objects not useable in all of the contexts where the traditional "first-class citizens" of a DBMS (integers, character strings, etc.) may appear. Here are some examples where rethinking the way we handle certain elements could improve the usability of a system.

Object Oriented (OO) and Object-relational (OR) DBMSs showed how text and other data types can be added to a DBMS and how to extend the query language with functions that operate on these extended data types. Current database systems have taken their first steps toward supporting queries across text and structured data; but they are still inadequate for integrating structured data retrieval with the probabilistic-reasoning characteristic of information retrieval. To do better, we need a fresh approach to both data models and the query language required to access both text and data. At the very least, probabilistic reasoning and other techniques for managing uncertainty must become first-class citizens of the DBMS.

Likewise, a major addition to recent DBMSs is their ability to add user-defined procedures to the query language. This approach allows one to add a new data type along with its behavior (methods). Unfortunately, this approach makes procedures second class citizens. We would like to see code become a first-class citizen of the DBMS, as well.

Triggers and active databases are another source of executable code inside a DBMS. Often, one wants to be alerted if a specific database condition becomes satisfied or a certain event occurs. If there are millions of such conditions, it is inefficient or even infeasible to poll the database periodically, to see which of the conditions are true. Rather, one wants to specify the monitoring condition to the DBMS and then have the DBMS alert the user asynchronously if the indicated condition becomes true. Commercial vendors have added triggers and alerters to their products, and there has been considerable research on how to make such facilities scalable. However, triggers and alerters have been grafted onto existing DBMS architectures. While it is not feasible to reason completely about code in the general case, it would be useful to have ways for the DBMS to do simple, perhaps only syntactic, reasoning about code objects. For instance, we could hope for the ability to find all code that depends upon a given database object.

We expect that several emerging application classes will force data streams to become a first-class part of the DBMS as well. The imminent arrival of commercial microsensor devices at low cost will enable new classes of "monitoring" DBMS applications. It will become practical to tag every object of importance with a sensor that will report its state in real time. For example, instead of attaching a property tag to items such as laptop computers and projectors, one will attach a sensor. In this case, one can query a monitoring system for the location of a lost or stolen projector. Such monitoring applications will be fed "streams" of sensor information on objects of interest. Such streams will put new demands on DBMSs in the areas of high performance data input, time-series functionality, maintenance of histories, and efficient queue processing. Presumably, commercial DBMSs will try to support monitoring applications by grafting stream processing onto the traditional structured-data architecture.

Lastly, there is a new form of science emerging. Each scientific discipline is generating huge data volumes, for example, from accelerators (physics), telescopes (astronomy), remote sensors (earth sciences), and DNA microarrays (biology). Simulations are also generating massive datasets. Organizing, analyzing and summarizing these huge scientific datasets stands as a real DBMS challenge. So is the positioning and transfer of data among various processing and analysis routines distributed through a grid, which requires knowledge of the overall structure of the processing chain and the needs and behavior of each module in it. It will require an integration of data and procedures that allows complex objects and advanced data analysis to be integrated with the DBMS.

In our opinion, it is time to stop grafting new constructs onto the traditional architecture of the past. Instead, we should rethink basic DBMS architecture with an eye toward supporting:
- Structured data
- Text, space, time, image, and multimedia data
- Procedural data, that is data types and the methods that encapsulate them
- Triggers
- Data streams and queues

as co-equal first-class components within the DBMS architecture — both its interface and its implementation — rather than as afterthoughts grafted onto a relational core.

The participants suggested that it would be better for the research community to start with a clean sheet of paper in many cases. Attempts to add these capabilities to, SQL, XML Schema and XQuery, are likely to result in unwieldy systems that lack a coherent core. Because of their forced dependence on prior standards, we believe strongly that XML Schema and XQuery are already too complex to be the basis for this sort of new architecture. The self-describing record format is a great idea for communication of information, but it is not especially convenient for the DBMS we envision, where procedures, text, and structured data are co-equal participants. Lastly, a new information architecture cannot be burdened by the political compromises of the past. We believe that vendors will pursue the extend-SQL and extend-XML strategies, to improve their existing products incrementally. By contrast, the research community should explore a reconceptualization of the problem.

A start on such an architecture should be a five-year goal for our community. As a concrete milestone, we look for several substantial prototypes before the next meeting of our ad-hoc group.

### 3.2 Information Fusion

Enterprises have tackled information integration inside their semantic domains for more than a decade. The typical approach is to extract operational data, transform the resulting data into a common schema, and then load the result into a data warehouse for subsequent querying. In this mode, information integration is performed in advance, typically by an extract-transform-load (ETL) tool to build the data warehouse and data marts. This is a feasible approach within an enterprise with a few dozen operational systems under the control of a single corporation.

The Internet completely breaks this extract-transform-load paradigm. There is now a need to perform information integration among different enterprises, often on an ad-hoc basis. Few organizations will allow outside entities to extract all of the data from their operational systems, so the data must stay at the sources and be accessed at query time. Some commercial products do this task today, but with a relatively-small, static set of sources within one enterprise.

As mentioned earlier, sensor networks and the new science will be generating huge datasets. These sensors and datasets will be distributed throughout the world, and can come and go dynamically. This breaks the traditional information integration paradigm, since there is no practical way to apply an ETL tool to each such occurrence.

Therefore, one must perform information integration on-the-fly over perhaps millions of information sources. The DBMS research community has investigated federated data systems for many years. The first of these reports [1] talked extensively about the problem. However, the thorny question of semantic heterogeneity remains. Any two schemas that were designed by different people will never be identical. They will have different units (your salary is in Euros, mine is in dollars), different semantic interpretations (your salary is net including a lunch allowance, mine is gross), and different names for the same thing (Samuel Clemens is in your database but Mark Twain is in mine). A semantic-heterogeneity solution capable of deployment at web scale remains elusive. Our community must seriously focus on this issue, or cross-enterprise information integration will remain a pipe dream. The same problem is being investigated in the context of the semantic Web. Collaboration by groups working on these and other related problems, both inside and outside the database community, is important.

There are many other difficult problems to be solved before effective web-scale information integration becomes a reality. For example, current federated query execution systems send subqueries to every site that might have data relevant to answer a query, thereby giving a complete answer to every query. At web scale, this is infeasible and query execution must move to a probabilistic world of evidence accumulation and away from exact answers. As another example, conventional information integration tacitly assumes that the information in each database can be freely shared. When information systems span autonomous enterprises, query processing must be done such that each database reveals only the minimal information necessary to answer the query and in conformance with its security policy. A third example is tying information integration to monitoring applications that span multiple data sources. For example, let me know when any of my mileage plans is giving bonuses on hotel stays for chains that have hotels near the sites of conferences or meetings I will be attending.

### 3.3 Sensor Data and Sensor Networks

Sensor networks consist of very large numbers of low-cost devices, each of which is a data source, measuring some quantity: the object's location, or the ambient temperature, for example. We mentioned before that these networks provide important data sources and create new data-management requirements. For instance, these devices will generally be self-powered, wireless devices. Such a device draws far more power when communicating than when computing. Thus, when querying the information in the network as a whole, it will often be preferable to distribute as much of the computation as possible to the individual nodes. In effect, the network becomes a new kind of database machine, whose optimal use requires operations to be pushed as close to the data as possible.

Query execution on sensor networks requires a new capacity: the ability to adapt to rapidly changing configurations, such as sensors that die or disconnect from the network. The query plan needs to change as the network changes, a capability we do not see among database systems today.

Sensors also suggest the need to deal with more complex forms of information integration. A common case is when sensors are not completely calibrated. A value from a sensor needs to be interpreted in the light of what other sensors are saying. A more complex matter is that the goal of sensor-data processing may be to deduce a very high-level fact from very low-level signals. For instance, we may want to combine heat, sound, and vibration sensors to locate a person nearby.

### 3.4 Multimedia Queries

Obviously, the amount of multimedia data (pictures, video, audio, etc.) is increasing dramatically. A challenge to our community is to create easy ways to analyze, summarize, search, and view the "electronic shoebox" of a person's multimedia information. The challenge ranges from Vannevar Bush's Memex [8] vision to the prosaic task of preparing a multimedia presentation about the children for Aunt Betsy. Any of these goals requires much better facilities for managing multimedia information than are available today.

### 3.5 Reasoning about Uncertain Data

Traditional DBMSs were applied to business data processing, which typically focused on numbers and character strings. In those application areas, data elements are precise quantities like address, quantity on hand, balance, status, and delivery date. As a result, current DBMSs have no facilities for either approximate data or imprecise queries.

When one leaves business data processing, essentially all data is uncertain or imprecise. Scientific measurements have standard errors. Location data for moving objects involves uncertainty in current position. Sequence, image, and text similarity are approximate metrics. To analyze imprecision, the "lineage" (or "provenance") of the data must be tracked, since a scientist needs to know where the data came from (the instruments, the settings of those instruments) and what cleaning, rescaling, remodeling, etc. was done subsequently to arrive at the data to be interpreted. Clearly DBMSs need built-in support for data imprecision.

As noted in previous sections, query processing must move from a deterministic model, where there is an exact answer for every query, to a stochastic one, where the query processor performs evidence accumulation to get a better and better answer to a user query. Users should also be able to ask imprecise queries and have the processing engine include this further source of uncertainty. Of course, with imprecise answers comes a duty for the system to characterize the accuracy offered, so users can understand whether or not the approximation is good enough for their needs. For instance, information retrieval systems measure precision and recall to help researchers understand how good an answer is.

### 3.6 Personalization

Several participants noted that query answers should depend on the profile of the person who is asking. The answer for a domain expert should be different from the answer for a novice. Relevance and relevance feedback should also depend on the person and the context. It should be possible for data at multiple sources that is organized for one purpose to be offered for other purposes. For example, health information that is organized for health-care providers should be personalized for individual use (e.g. hospital records, prescriptions, drug interaction information, family medical history, immunization records, dental records, and insurance claims). Mass personalization should be feasible in next-generation information systems. What is needed is a framework for including and exploiting appropriate metadata for such personalization.

Participants also noted that personalization and uncertainty leave one with a need to verify that the information system is producing a "correct" answer. For example, what happens if the information system is buggy and produces the wrong approximate or personalized answer?

### 3.7 Data Mining

Historically, data mining has focused on efficient ways to discover models of existing data sets. These models must expose some useful aspect of the data, while obscuring details that are not useful for the intended application. Algorithms have been developed by many research communities to perform such operations as classification, clustering, association-rule discovery, and summarization. These techniques are now part of mainstream products from the major DBMS vendors. Today data mining and business intelligence is heavily used by Fortune 500 end users and by small applications alike. The success of current data mining tools has created a hunger for the next generation of tools: Fortune-500 warehouse users invariably point out that they have a single data mining query: "Tell me something interesting." They are glad to have the current crop of data mining tools, but they wish for tools that are better at generating these unexpected "pearls of wisdom."

A challenge for data-mining research is to develop algorithms and structures for sifting through the database looking for such pearls, while running in the background and consuming excess system resources. Another important challenge is to integrate data mining with querying, optimization, and other database facilities such as triggers. We hope that data mining moves in this direction and beyond algorithms for elementary operations. There was also a feeling that computer science and IT curricula should include more about the proper use of data-mining tools.

### 3.8 Self Adaptation

One of the consequences of the widespread use of DBMS technology is a shortage of competent database administrators. Modern DBMSs are much more complex than their counterparts of 20 years ago. Today, a database administrator (DBA) must understand disk partitioning, parallel query execution, thread pools, and user-defined data types. These concepts were not present in the systems of yesteryear. Put succinctly, current DBMSs are thought to be too hard to use. To address this shortcoming, the major vendors have all embarked on projects to simplify database administration.

Such projects have at least two components. First, current DBMSs have a large collection of tuning knobs. These allow an expert to extract the best performance from an operational system. Often, such tuning is done by the vendor's engineers, at great expense to the customer. Most systems engineers doing this tuning do not, in fact, have a deep understanding of the meaning of the knobs. Instead, they have seen many system configurations and workloads and carry a "crib sheet" of tuning parameters that had optimized other systems. When presented with a new environment, they reach for the crib sheet that is closest to the configuration at hand and use those settings.

It should be possible to perform tuning using a combination of a rule-based system and a database of knob settings and configuration data. There has been substantial progress in this direction, mostly by the vendors, in such areas as dynamic resource allocation, the selection of physical structures and, to some extent the selection of materialized views (redundant data maintained by the DBMS to speed up the execution of certain queries). In our opinion, the ultimate goal is "no knobs," where all tuning decisions are made automatically by the system, perhaps guided by default policies, such as the relative importance of response time and throughput, or by personal profiles that summarize user needs. Thus, more sophisticated models of user behaviors and workloads are a prerequisite for progress in this area. We believe it should be possible to achieve real no-knobs operation and recommend this goal as a major focus of the research community.

Many new applications that use DBMSs are going to require unattended operation. In addition to no-knobs tuning, the DBMS must be able to recognize internal malfunctions and malfunctions of communicating components, identify data corruption, detect application failures, and do something about them. Such capabilities require making the DBMS more self-aware and providing it with explicit models of the information system in which it participates.

### 3.9 Privacy

The widespread adoption of the Internet has created a surge in information available about individuals. Moreover, there is plenty of horsepower available to perform correlations among databases. This convergence has led to an unprecedented amount of information that can be discovered about individuals. Mundane data, such as every address at which one has ever lived, is readily obtained. It appears easy to discover every person who has ever lived at a given address, making it easy to discover all the neighbors of any given person ever had. It is claimed that persons who rode on the same airplane flight can also be obtained. Furthermore, identity theft is a troubling national problem, since it is not hard to obtain vital records of deceased persons as well as information needed to apply for a credit card in a phony name.

Data-oriented security research was prevalent in the 1980's but has died down since then. We see a need to revitalize this subfield, but with a distinctly different orientation. Today, we need to address the concerns, policies and mechanisms to support multiple individual options and controls on information held by third parties. This collection of issues is likely to be quite different from those of the pre-Web security model of a single organization protecting its data from. While laws will continue to have an important role in addressing information privacy and related security issues, we can change the set of options available and reach better points in the space of privacy-security tradeoffs by advancing what is technically realizable.

Since much of this information correlation is performed by DBMSs, our community can work on security systems that include a component dealing with the prospective use to which the data will be put. Access decisions should be based not only on who is requesting the data but also on to what use it will be put. Moreover, declarative systems that specify the purpose of the data request are something our community should be good at, since we have dealt with data-oriented declarative specifications for in other contexts.

**3.10 Trustworthy Systems**

Privacy is only one aspect of the broader issue of trustworthy systems that safely store data, protect it from unauthorized disclosure, protect it from loss, and make it always available to authorized users. There is increasing interest in digital rights management — protecting intellectual property rights and allowing private conversations. Furthermore, ensuring the correctness of query results and data-intensive computations is a growing concern, especially for embedded systems such as health-care smart cards and other medical applications. Logical inferencing technology may be helpful in validating correctness, such as theorem proving and model checking. The information-management community should play a central role in addressing these needs and enhancing DBMSs with mechanisms to support these capabilities.

**3.11 New User Interfaces**

It has been a common lament throughout the years that the database community does too little about user interfaces. There is now the horsepower on the desktop to run very sophisticated visualization systems. However, for a given type of information coming from a DBMS, it is not clear how best to render it visually. A small number of slick visualization systems oriented toward information presentation were proposed during the 1980's, notably QBE and VisiCalc. There have not been comparable advances in the last 15 years, and there is a crying need for better ideas in this area.

Thirty years of research on query languages can be summarized by: "we have moved from SQL to XQuery." At best, we have moved from one declarative language to a second declarative language with roughly the same level of expressiveness. It has been well documented that end users will not learn SQL; rather SQL is a notation for professional programmers. We see in other communities a number of ideas that could impact database-oriented research on interfaces. The information-retrieval community has used keyword-based querying for decades. Browsing has become increasingly popular in a number of areas.

Perhaps most interesting is the research opportunities suggested by the term "semantic Web." While it may be unclear what the concept truly entails, much of the recent work has centered on "ontologies." An ontology characterizes a field or domain of discourse by identifying concepts and relationships between them, usually in a formal language. We mentioned in Section 2.2 how this work may support information integration, since a fundamental problem in that area is the inability to combine databases that at a deep level are talking about the same thing, but do so in different terminology. Work on ontologies may likewise enable users of databases and other resources to use speech or natural language to query in their own terminology. The database community should be looking for opportunities to exploit these developments in future database management systems.

**3.12 One-Hundred-Year Storage**

As an ever-increasing fraction of the world's information is stored digitally, there is a need for indefinite electronic storage of information. However, even archived information is disappearing, because it was captured on a medium that is deteriorating (e.g. photographic film or magnetic tape), because it was captured on a medium that requires obsolete devices (e.g. special storage drives), or because the application that is needed to interpret the information no longer works. Avoiding this information loss requires mechanisms for migration, to copy information from deteriorating or obsolete media, and for emulation, to capture methods that can interpret information that is stored for long periods.

Metadata also plays an important role. For example, the capture of scientific data requires that the "lineage" of the data also be captured as well as any reader programs that are required to gain access to the data. It may also be essential to have metadata describing context. For example, tables on social service spending in Germany over the years might not explicitly show that the 1983 tables are in Deutschmarks while the 2003 tables are in Euros, or that the 1983 table only included states in West Germany. Without such implicit context, such data is worthless, even if it is accessible. In general, society would benefit immensely by our information management research community architecting such a store, whose content remains accessible in a useful form, indefinitely. Such a store would, to the extent possible, automate the process of migrating content between formats or maintaining the hardware and software that each document needs to be used. It would also manage the metadata along with the stored documents themselves.

**3.13 Query Optimization**

Many of the participants saw query optimization as an important element of one or another of the endeavors discussed above. The general principle is that when we deal with very large volumes of data, we tend to manipulate that data in a regular way. This regularity allows very-high-level languages like SQL or XQuery to be used in the database world successfully, but almost nowhere else. But very-high-level languages demand a competent optimizer. Participants suggested that we need further work on optimization of information integrators, for semistructured query languages like XQuery, for stream processors, for sensor networks, and possibly for other domains as well

It was also observed that many uses of SQL systems involve sequences of relatively simple queries, embedded in a host-language program, all cooperating to accomplish one task. We see the need to consider inter-query optimization involving large numbers of queries, even in the very traditional, pure-relational setting.

# 4 Next Steps and Discussion

Many of the research directions suggested in the body of the report are very long-term goals. However, the Lowell participants had several suggestions for next steps that should be readily accomplishable by the time the next meeting is held, five or so years from now. Several goals were mentioned in the narrative above, such as the proposal to reconsider DBMS architecture to handle new data types, approximate reasoning, and to treat both procedures and data as co-equal. In this section, we call out some other next steps.

We think that information-integration research would be well served by generating a test bed and collection of integration tasks. Doing so would allow easy access to a test platform for anybody with an integration idea. The test bed would allow a controlled way to compare solutions, including the opportunity to gain bragging rights, if one is currently the best at one or another of these integration tasks. And it would help generate interest in this research area. Several researchers pointed out that TREC [7] has served the same purpose in the IR community

There was considerable discussion about how to construct such a test bed. While there are many design issues to enable controlled experiments, the most pressing problem is obtaining appropriate data sets. One potentially practical proposal was for 10--20 Computer Science departments to make public some of their nonproprietary classroom scheduling or other similar data. Any member of a CS department who can make such a data set available is urged to contact Mike Stonebraker, who will coordinate the construction of this test bed. Perhaps a more sophisticated test bed with much larger data sets can be proposed.

There was considerable support for the position that many of the traditional research focuses were solved problems (e.g. ACID properties). Others countered that "sea changes" in related technology might require revisiting these problems, and that there is always the potential of finding simpler mechanisms that can be more easily implemented or more powerful mechanisms that can be more broadly applied. There was also a heated discussion concerning whether stream-processing systems would require a new DBMS engine or whether current ones could adapt to the new requirements successfully.

Discussion also centered on the level at which information integration should occur. Although many thought the DBMS was the best place to do it, others thought it would be more natural and widespread at the application (i.e., web services) level. There was also considerable discussion about whether Web services would make any progress on addressing semantic-heterogeneity issues. Some thought that de-facto standards would emerge for such services; while others countered that the electronic component community has been trying hard for years to standardize on a set of services (Rosetta Net) with limited success.

We close this report with two observations on which there was general agreement. First, the database research community should avoid drawing too narrow a box around what we do. We need to explore opportunities for combining database and related technologies that can improve the usage of information, such as information visualization technologies, which has often been left to the domain of other research communities. To broaden the set of technologies database researchers apply, they need to expand their breadth of competencies. One is reminded of the plasterers union which decided many years ago, when wallboard was being introduced, that it was not their competency. As plaster was replaced by wallboard, the union lost out.  This fate could befall the DBMS community if it does not respond to the new challenges of integrating related technologies with information management.

Second, it was noted that the average age of participants at these meetings has been increasing.  On the other hand, there are more young database researchers than ever before, as evidenced by the large number of junior faculty in databases. We recommend that the next meeting should attempt to invite a broader mix of age-groups within our community.

# 5 References

1. Philip A. Bernstein, Umeshwar Dayal, David J. DeWitt, Dieter Gawlick, Jim Gray, Matthias Jarke, Bruce G. Lindsay, Pete C. Lockemann, David Maier, Erich J. Neuhold, Andreas Reuter, Lawrence A. Rowe, Hans-Jörg Schek, Joachim W. Schmidt, Michael Schrefl, and Michael Stonebraker: Future Directions in DBMS Research - The Laguna Beach Participants. SIGMOD Record 18(1): 17-26 (1989)

2. Abraham Silberschatz, Michael Stonebraker, and Jeffrey D. Ullman: Database Systems: Achievements and Opportunities. CACM 34(10): 110-120 (1991)

3. Abraham Silberschatz, Michael Stonebraker, and Jeffrey D. Ullman: Database Research; Achievements and Opportunities into the 21st Century. SIGMOD Record 25(1): 52-63 (1996)

4. Abraham Silberschatz, Stanley B. Zdonik, et al: Strategic Directions in Database Systems — Breaking Out of the Box. ACM Computing Surveys 28(4): 764-778 (Dec. 1996).

5. Philip A. Bernstein, Michael L. Brodie, Stefano Ceri, David J. DeWitt, Michael J. Franklin, Hector Garcia-Molina, Jim Gray, Gerald Held, Joseph M. Hellerstein, H. V. Jagadish, Michael Lesk, David Maier, Jeffrey F. Naughton, Hamid Pirahesh, Michael Stonebraker, and Jeffrey D. Ullman: The Asilomar Report on Database Research. SIGMOD Record 27(4): 74-80 (1998)

6. CRA Conference on "Grand Research Challenges" in Computer Science and Engineering, http://www.cra.org/Activities/grand.challenges/.

7. TREC Data home page, http://trec.nist.gov/data.html.

8. Vannevar Bush. "As We May Think." *Atlantic Monthly* (July 1945), pp. 101-108.